\documentclass[12pt]{article}

\usepackage{epsfig}

\begin{document}

\begin{center}
{\large\bf
COHERENCE AND SYNCHRONIZATION IN DIODE--LASER ARRAYS WITH DELAYED
GLOBAL COUPLING
}

\vskip5mm
{\large
Jordi Garc\'{\i}a--Ojalvo$^{1,2}$, Joan Casademont$^1$,
Claudio R. Mirasso$^3$,\\

M.C. Torrent$^1$ and J.M. Sancho$^4$
}

\vskip5mm
{\normalsize\em
$^1$ Departament de F\'{\i}sica i Enginyeria Nuclear, Universitat
Polit\`ecnica de Catalunya
\\Colom 11, E--08222 Terrassa, Spain

$^2$ Institut f\"ur Physik, Humboldt Universit\"at zu Berlin
\\Invalidenstr. 110, D--10115 Berlin, Germany

$^3$ Departament de F\'{\i}sica, Universitat de les Illes Balears
\\E--07071 Palma de Mallorca, Spain

$^4$ Departament d'Estructura i Constituents de la Mat\`eria, Universitat
de Barcelona
\\Diagonal 647, E--08028 Barcelona, Spain
}
\end{center}

\vskip5mm
\begin{abstract}
\noindent
The dynamics of a semiconductor--laser array whose individual elements
are coupled in a global way through an external mirror
is numerically analysed. A coherent in--phase solution is seen to be
preferred
by the system at intermediate values of the feedback coupling strength.
At low values of this parameter, a strong amplification of the
spontaneous emission noise is observed. A tendency towards chaos
synchronization is also observed at large values of the feedback strength.
\end{abstract}

\vskip1cm
\centerline{
RUNNING TITLE: Coherence and synchronization in laser arrays
}

\newpage

Semiconductor lasers are used in a wide range of applications including,
among other examples, optical communication systems, compact
disk and CD-ROM units, bar--code readers, and laser printers. Some of
their advantageous features are their small size, high efficiency,
low price, simple pumping requirements, and the possibility to modulate
their injection current at high frequencies. In general, semiconductor lasers
can emit either in a continuous or in a pulsating way, but in any case
the maximum emitted power is usually lower than a few tens of mW. However,
for many applications it is desirable to have high--power coherent
emission available from small sources. So far, edge--emitting
semiconductor lasers have been designed to produce relatively high
power (up to a few $W$) by arranging them in one--dimensional arrays
[Welch {\em et al.}, 1988].
More recently, two--dimensional arrays of vertical--cavity
surface--emitting lasers (VCSELs) have also been developed
[Sanders {\em et al.}, 1994].
Increase in the output power is obtained through either coherent 
or incoherent superposition of the laser beams emitted by the individual
elements in the array. In the coherent case, the phase difference between
neighboring beams has a constant value (the array is said to be {\em
phase locked}). This situation is usually preferred, since in this case
the maximum output power scales with the square of the number of elements
in the array. On the other hand, an incoherent (non phase--locked)
superposition leads to a total output power whose maximum value scales
linearly with the number of elements.

In order to have phase locking, some kind of coupling must exist between
the dynamics of the individual lasers of the array. In standard
diode--laser arrays, coupling between the elements occurs {\em locally}
through the smooth overlap of the evanescent field of each individual laser
with those of its nearest neighbours. In this case, the laser array can
reach a phase--locked state if the separation between neighboring elements 
is small enough (typically lower than 10 $\mu$m). Phase locking usually
arises in one of two ways: either all lasers emit with the same phase
(in--phase solution) or the phase difference between neighboring lasers
is fixed to a constant nonzero value (out--of--phase solution).
The in--phase solution
produces constructive interference in the optical axis, whereas for the
out--of--phase state, interference is destructive. Local coupling usually
leads to out--of--phase emission [Winful \& Wang, 1988], which is more
favorable energetically; in fact the stability range of the in--phase
solution is seen to be very small [Li \& Erneux, 1993].

In this work, we propose a different way to couple the elements of the
array, with the aim of favoring the in--phase coherent state.
Instead of using a local coupling like the one provided by the
evanescent field, we now consider a global, all--to--all coupling through
an external mirror at one side of the laser [Leger {\em et al.}, 1988]. 
Global coupling is known
to enlarge the stability range of the in--phase solution [Li \& Erneux,
1993]. With such a configuration, a coherent superposition of the
individual laser beams takes place at the mirror, and part of the resulting
electromagnetic field is fed back into the array, in such a way that each
individual laser is influenced by all other elements of the array in
an all--to--all coupling. A characteristic feature of this global coupling
is that it involves a delay, due to the time $\tau$ that the light takes
to propagate towards the mirror and back into the array. Other feedback
schemes have been recently
proposed to stabilize phase locking [Dasgupta \& Andersen, 1994] and
to control the chaotic output [Auerbach \& Yorke, 1996] of semiconductor
laser arrays. In those cases, however, the feedback loops are of
optoelectronic nature, whereas our current proposal involves all--optical
feedback coupling. {\em Local} feedback coupling has also been
recently proposed to control the transverse dynamics of broad--area
lasers [Martin--Regalado {\em et al.}, 1996].

In a first approximation, we ignore the evanescent local coupling
by considering a large enough separation between neighboring lasers of the
array. The array output is observed at the focal plane of a lens (at the
opposite side of the external mirror),
which is called the {\em far--field} plane and corresponds to the
Fourier transform of the system.

In order to analyse the behavior of the proposed scheme, we have
performed numerical simulations of a system of coupled rate--equations 
based on the Lang--Kobayashi model
[Lang \& Kobayashi, 1980]:
\begin{eqnarray}
& &\frac{d E_i}{d t}=\frac{1+i\alpha}{2}(G_i(E_i,N_i)-\gamma)\,E_i(t)
+\kappa e^{-i\omega\tau}\sum_{j=1}^n E_j(t-\tau)+\sqrt{2\beta N_i}
\xi_i(t)
\label{eq:model-E}
\\
\label{eq:model-N}
& &\frac{d N_i}{d t}=\gamma_e\,(CN_{th}-N_i)-G_i(E_i,N_i)\,|E_i(t)|^2
\end{eqnarray}
where $n$ is the number of lasers in the array, $E_i(t)$ is the complex
envelope of the electric field emitted by laser $i$, and $N_i$ is the
corresponding carrier number. The material gain is given by
$$
G_i(E,N)=\frac{g(N_i(t)-N_0)}{1+s|E_i(t)|^2}
$$
and the threshold value of the carrier number is $N_{th}=\gamma/g+N_0$.
Feedback is described by two parameters, the feedback strength $\kappa$
and the external round--trip time $\tau$. For simplicity, we have assumed that
both $\kappa$ and $\tau$ are the same for all the lasers of the array.
The feedback term also depends on the frequency $\omega=2\pi c/\lambda$,
where $\lambda$ is the wavelength of the emitted light and $c=3\times
10^8$ m/s is the speed of light in vacuum.
The random spontaneous emission process is
modeled by a complex Gaussian white noise term $\xi(t)$ of zero mean and
correlation $\langle\xi(t)\xi^*(t')\rangle =2\delta(t-t')$.
The definitions and values of the rest of parameters are given in Table 1.
Equations (\ref{eq:model-E})--(\ref{eq:model-N}) are numerically
integrated by means of a second--order Runge--Kutta algorithm for the
deterministic terms and a stochastic Euler method for the noise terms.

\begin{table}[htb]
\begin{center}
\begin{tabular}{|c|l|c|} 
\hline
Parameter & Description & Value \\
\hline
$C$ & relative bias current & $2.0$\\
$\gamma_e$ & inverse carrier lifetime & $5\times 10^8$ s$^{-1}$\\
$\gamma$ & inverse photon lifetime & $5\times 10^{11}$ s$^{-1}$\\
$\alpha$ & linewidth enhancement parameter & 5\\
$\lambda$ & laser wavelength & 1.5 $\mu$m\\
$g$ & differential gain coefficient & $1.5\times 10^4$ s$^{-1}$\\
$N_0$ & transparency inversion & 1.5$\times 10^8$ \\
$s$ & saturation coefficient & $10^{-6}$ \\
$\beta$ & noise intensity & $10^{3}$ s$^{-1}$ \\
\hline
\end{tabular}
\caption{Parameters of the diode--laser array model.}
\end{center}
\end{table}

In order to characterize the behavior of the system, we define
an incoherent intensity
\begin{equation}
S=\sum_{i=1}^n |E_i(t)|^2
\label{eq:incint-def}
\end{equation}
and a coherent intensity
\begin{equation}
I=\left|\sum_{i=1}^n E_i(t)\right|^2
\label{eq:cint-def}
\end{equation}
The physical interpretation of these magnitudes is the following. $S$
is the sum of the individual laser intensities; it can be
measured by placing a broad--area detector next to the output face of
the array, and it is defined in such a way that $S/n$ gives the output
level of emission from one individual laser. The
intensity $I$, on the other hand, corresponds to a coherent superposition
of the individual electric fields; it can be measured by placing a
detector at the far--field plane of the lens.

\begin{figure}[htb]
\begin{center}
\epsfig{file=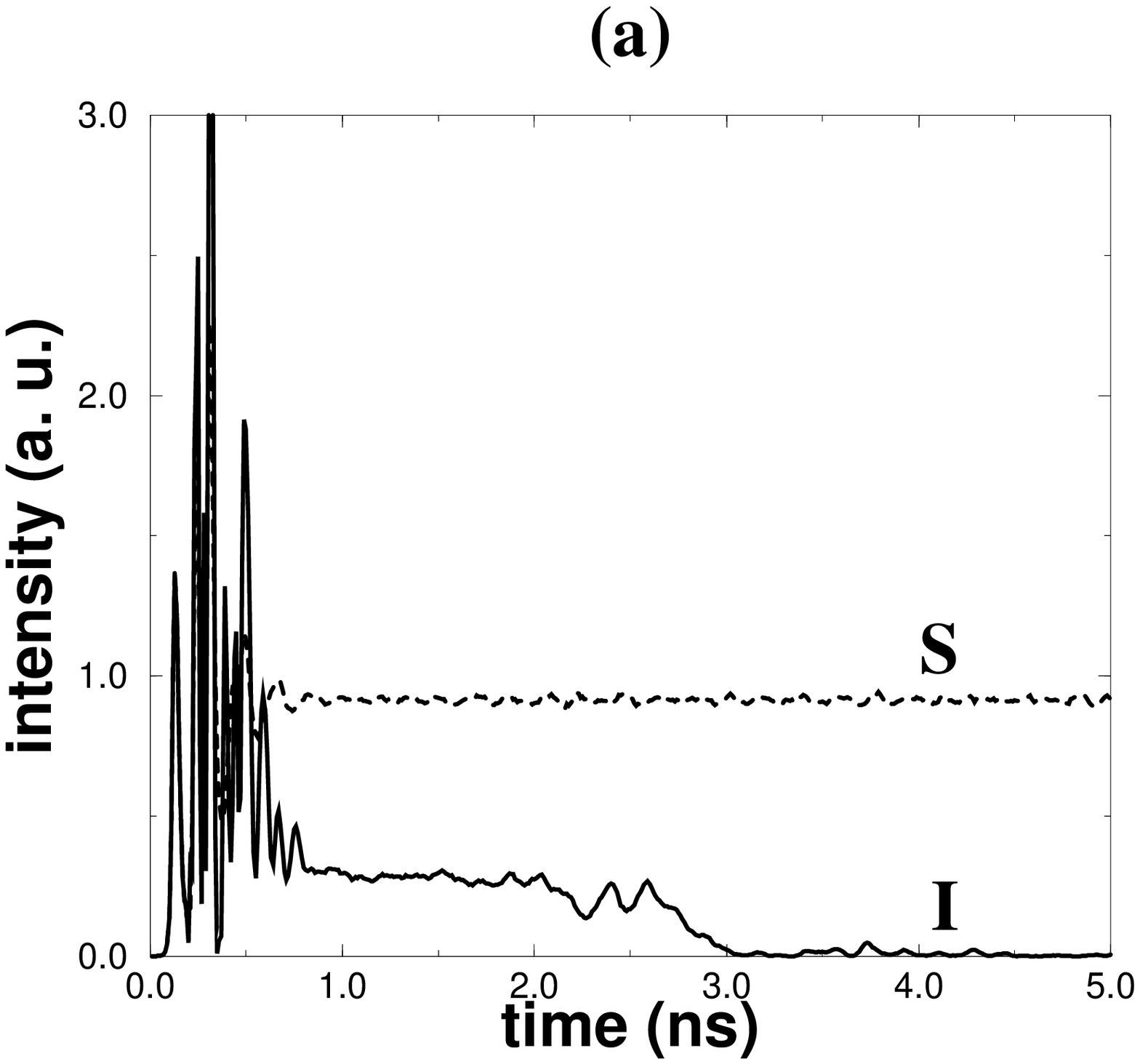, width=3.0in}
\epsfig{file=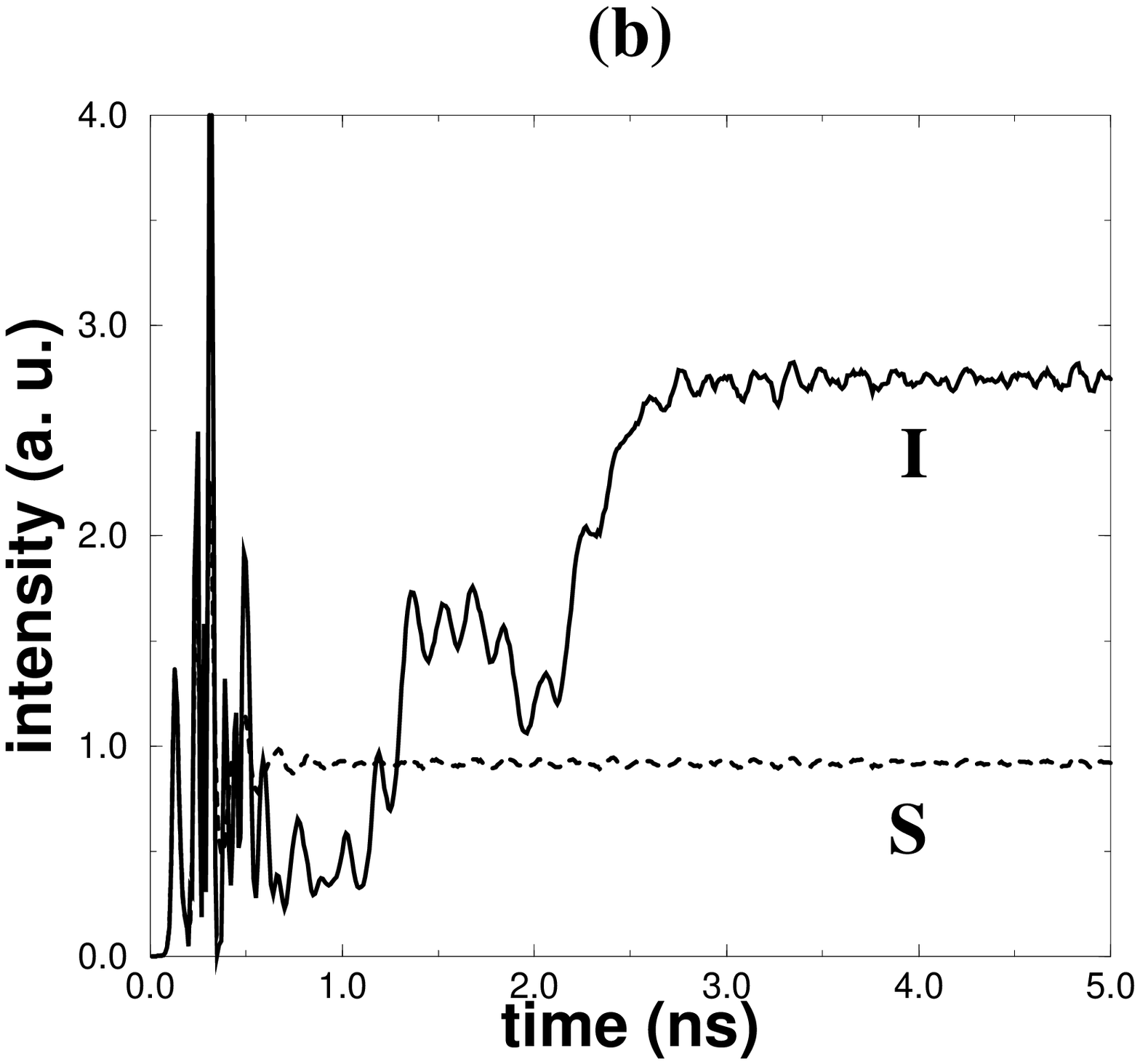, width=3.0in}
\caption{Phase locked behavior of the diode--laser array for
$\tau=500$ ps.
(a) Out--of--phase solution for $\kappa=10^{-4}$ ps$^{-1}$;
(b) in--phase solution for $\kappa=9\times 10^{-4}$ ps$^{-1}$.}
\label{fig:locked}
\end{center}
\end{figure}

Typical phase--locked states of the system are shown in Fig.
\ref{fig:locked}, for a three--laser array and a delay time
$\tau=500$ ps. At low values of the feedback strength $\kappa$
an out--of--phase solution arises (Fig. \ref{fig:locked}a),
characterized by a value of $I$ close to 0 (destructive interference); 
in fact the intensity profile
in the far field has in this case a double--lobed structure with
a minimum at the optical axis. We have also observed that
the phase difference between neighboring lasers is $2\pi/3$.

Contrarily to what happens in locally coupled arrays,
an in--phase solution can now be easily obtained by increasing
the feedback strength. Figure \ref{fig:locked}b shows such an
in--phase behavior, for which $I\approx nS$ (which means that
the intensity at the far--field plane 
is $n^2$ times the output from a single laser);
the intensity profile in the far field is now single lobed.
The phase difference between neighboring lasers is 0. 
We have checked that in most of the cases we tested, different values
of the $\kappa$ parameter for different lasers of the array
still give in--phase solution, although the
output power can be lower than in the case of equal $\kappa$.
A detailed study of the synchronization dependence on feedback
parameters is left for a near future.
We should
note that the range of stability of this in--phase solution is
very broad; a transition from the out--of--phase to the in--phase
state can be easily induced simply by increasing $\kappa$.
Preliminary results show that this transition is hysteretical,
which indicates the possibility of coexistence between the
in--phase and out--of--phase coherent states.
The stability region of the in--phase state
depends on the external round--trip time $\tau$.
In particular,
simulations show that for values of $\tau$ not too
large, the width of this region increases when $\tau$ decreases.
Additionally, and according to what is known in the case of a single
semiconductor laser with optical feedback, the phase accumulated
by the field in the external cavity, $\phi_{ext}=\omega \tau$,
should also play an important role in the behavior of the system.
For the parameters that we have chosen and
for $\tau=500$ ps, this accumulated phase is $\phi_{ext}$=0
(mod$_{2 \pi}$). We have checked that up to values $\phi_{ext}
\,\sim\,\pm 0.5$ (mod$_{2 \pi}$), in--phase coherence is still
well reached. A more detailed study of this effect will be carried
out in the future.

\begin{figure}[htb]
\begin{center}
\epsfig{file=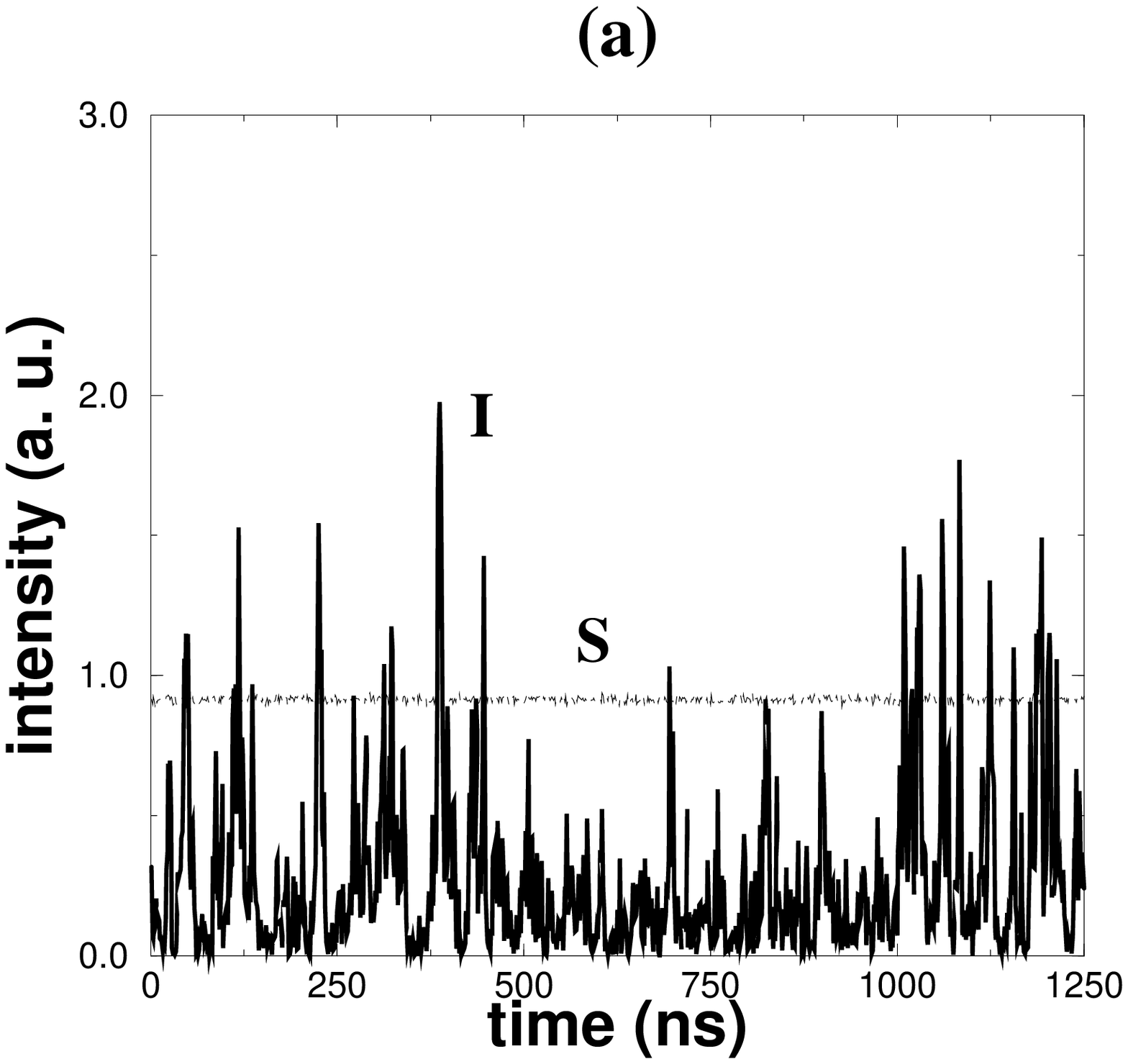, width=3.0in}
\epsfig{file=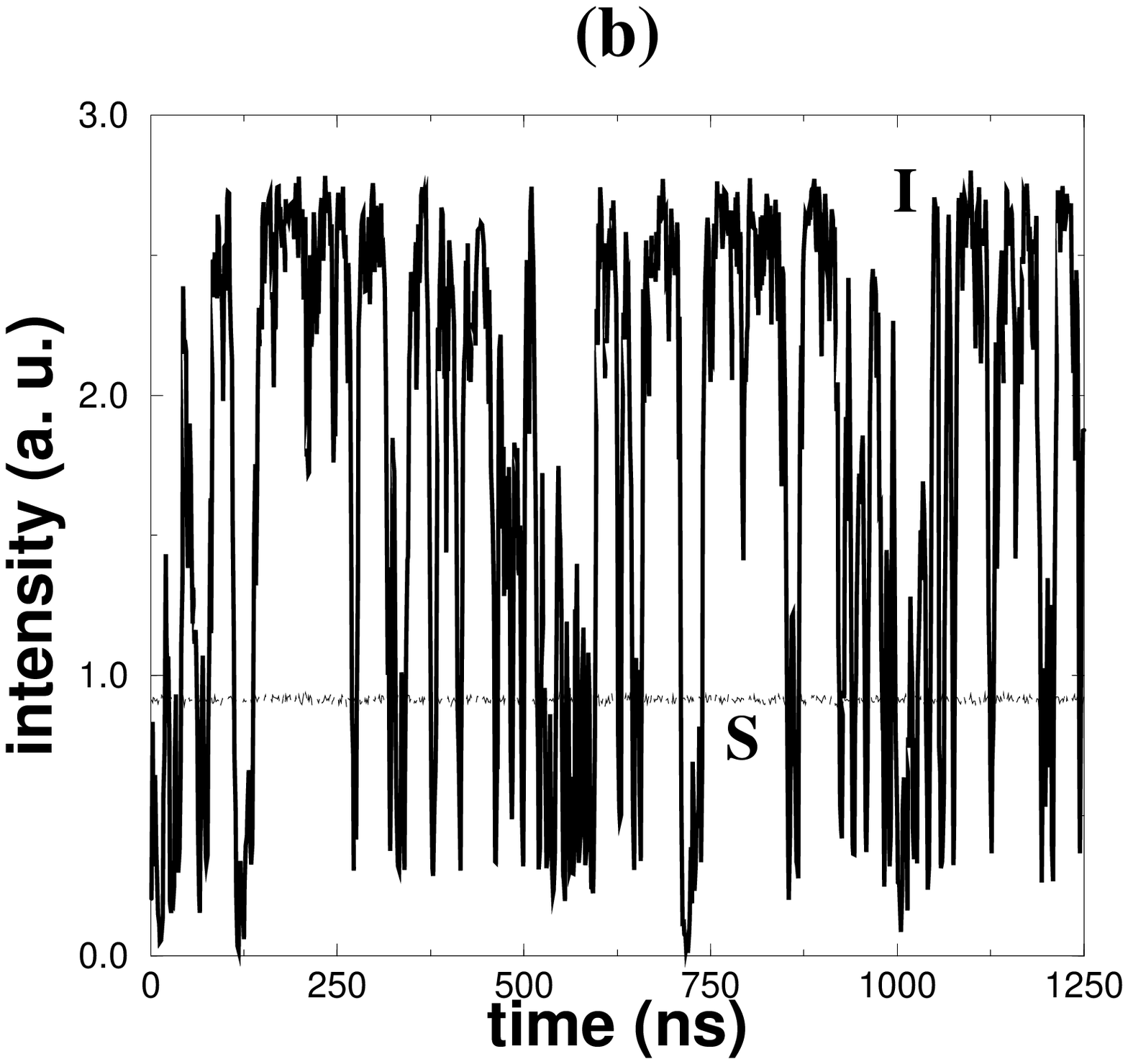, width=3.0in}
\caption{Incoherent bursts of the dynamics for $\kappa=5\times
10^{-5}$ ps$^{-1}$.
(a) Departure from the out--of--phase state for $\tau=250$ ps;
(b) fluctuating jumps between in--phase and out--of--phase behavior
for $\tau=125$ ps. This phenomenon disappears if spontaneous--emission
noise is not considered in the model.}
\label{fig:amp}
\end{center}
\end{figure}

For small values of the feedback strength $\kappa$, 
the delay time $\tau$ starts to have an important role in the
dynamics of the system. For $\tau$ large enough ($\tau\sim 1$ ns)
the system develops an out--of--phase coherent state such as
the one shown in Fig. \ref{fig:locked}a. However, when $\tau$
decreases bursts of incoherent behavior start to appear, leading
to a highly fluctuating value of the coherent intensity $I$.
This situation is shown in Fig. \ref{fig:amp} for
$\kappa=5\times 10^{-5}$ ps$^{-1}$ and two small values of the
delay time $\tau$. For $\tau=250$ ps (Fig. \ref{fig:amp}a),
high--amplitude bursts of incoherent behavior appear from the
out--of--phase state. This highly fluctuating dynamics intensifies
for smaller $\tau$: for $\tau=125\,ps$ (Fig. \ref{fig:amp}b),
the system jumps wildly between the in--phase and out--of--phase
states.
A na\"{\i}ve interpretation of this behavior can be given by
reminding that a decreasing delay time tends to increase the stability
range of the in--phase solution. This tendency contrasts with the
destabilizing role of a decreasing feedback strength. A competition
between these two effects could be the reason for the highly
fluctuating behavior of the system. It should also be noted that
this complex phenomenon only occurs if spontaneous emission
noise (as described by the stochastic term $\xi_i(t)$ in
Eq. (\ref{eq:model-E})) is considered. Hence it can be said that
one is observing an intense noise amplification for low values
of $\kappa$ and $\tau$. 

\begin{figure}[htb]
\begin{center}
\epsfig{file=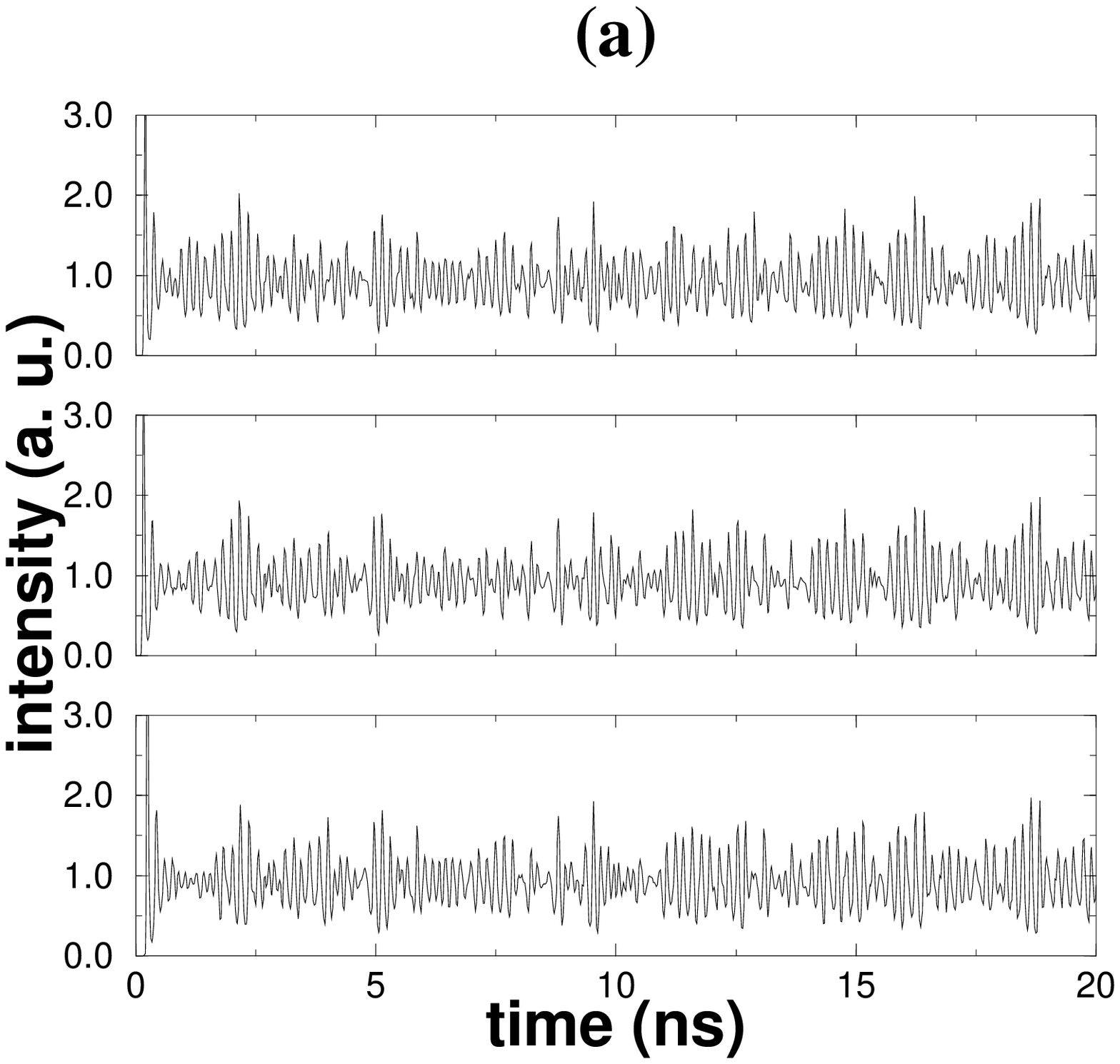, width=3.0in}
\epsfig{file=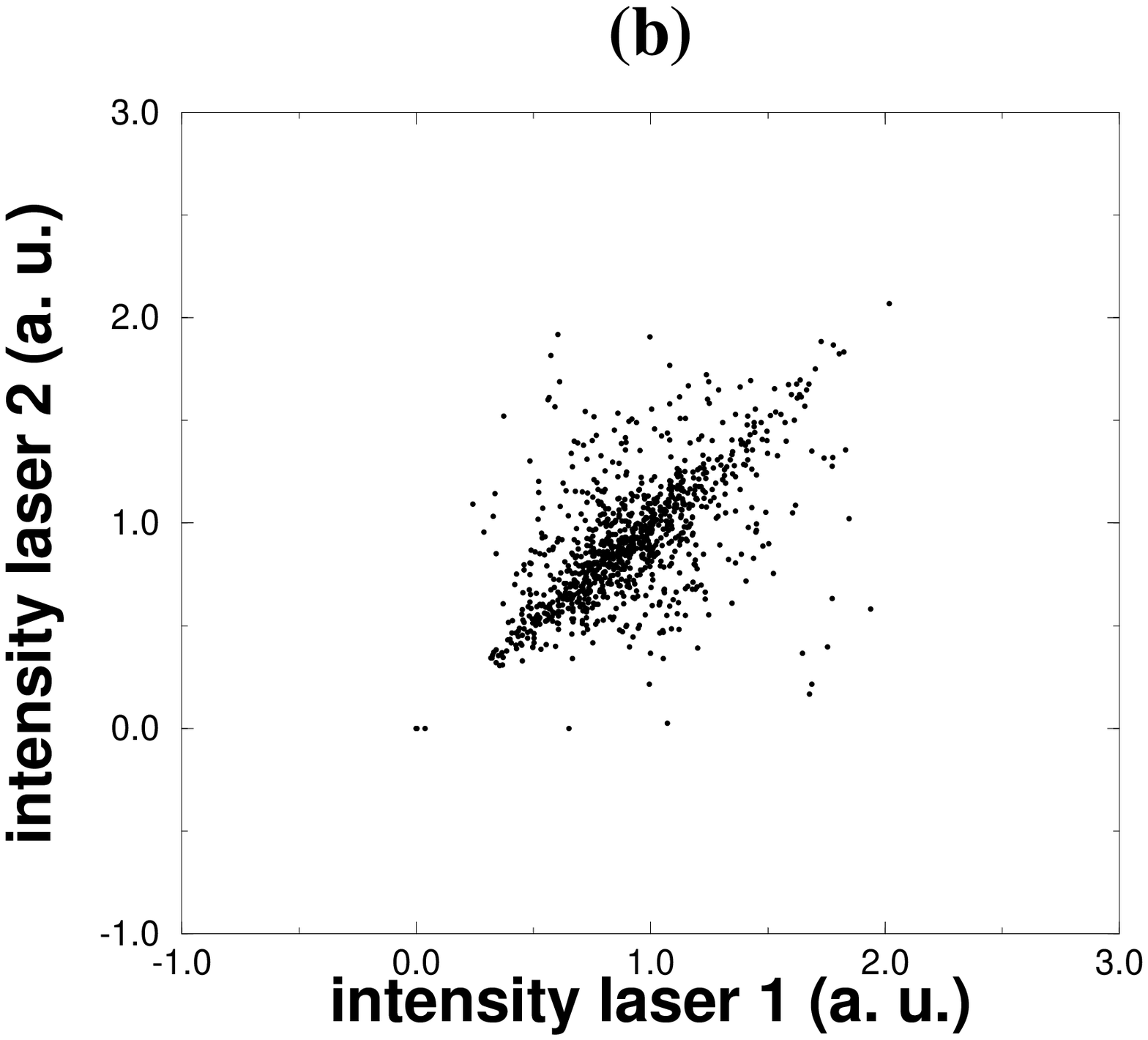, width=3.0in}
\caption{Chaotic synchronization in a three--laser array for
large feedback strength $\kappa=0.002$ ps$^{-1}$ and $\tau=500$ ps.
(a) Chaotic oscillations of the individual lasers;
(b) synchronization plot of laser 1 vs. laser 2.}
\label{fig:sync}
\end{center}
\end{figure}

Finally, for large values of the feedback coupling strength
$\kappa$, the array undergoes chaotic oscillations, as shown in
Fig. \ref{fig:sync} for a three--laser array with $\kappa=0.002$
ps$^{-1}$ and $\tau=500$ ps. This situation is to be expected,
since single diode lasers exhibit a rich chaotic dynamics in the
presence of optical feedback [Sacher {\em et al.}, 1992].
Indeed, each individual laser of the array behaves chaotically,
as shown in Fig. \ref{fig:sync}a, whereas some tendency towards
synchronization can be observed by plotting the output intensity
of one of the lasers versus another (Fig. \ref{fig:sync}b).
Chaos synchronization in laser arrays was predicted numerically
by Winful \& Rahman [1990] and observed experimentally by 
Roy \& Thornburg [1994].

In conclusion, we have numerically analysed the dynamical
behavior of a semiconductor laser array in which the individual
elements are coupled in a {\em global} way through an optical feedback. 
The system is seen to select an in--phase coherent
solution for intermediate values of the feedback coupling
strength $\kappa$. For low values of $\kappa$, noise amplification
is observed in the form of random bursts of incoherent behavior.
Finally, for large values of $\kappa$ the dynamics of the
individual lasers is chaotic, and some tendency towards
chaos synchronization is observed.

We thank R. Roy and R. Vilaseca for fruitful discussions. This
research is supported by the Direcci\'on General de Investigaci\'on
Cient\'{\i}fica y T\'ecnica (Spain), through project PB96--0241,
and by the Comisi\'on Interdepartamental de Ciencia y Tecnolog\'{\i}a
(Spain), through project TIC97--0420. Computing support is acknowledged
from the Centre de Computaci\'o i Comunicacions de Catalunya (C$^4$).
J.G.O. also acknowledges financial support from the Alexander
von Humboldt--Stiftung (Germany).
\pagebreak

{\large\bf REFERENCES}

\begin{itemize}
\item[] Auerbach, D. \& Yorke, J.A. [1996] "Controlling chaotic
fluctuations in semiconductor laser arrays," {\em J. Opt. Soc. Am.
B} {\bf 13}(10), 2178--2187.
\item[] Dasgupta, S. \& Andersen D.R. [1994] "Feedback stabilization
of semiconductor laser arrays," {\em J. Opt. Soc. Am. B} {\bf 11}(2),
290--296.
\item[] Lang, R. \& Kobayashi, K. [1980] "External optical feedback
effects on semiconductor injection laser properties,"
{\em IEEE J. Quantum Electron.} {\bf QE--16}(3), 347--355.
\item[] Leger, J.R., Scott, M.L. \& Veldkamp, W.B. [1988]
"Coherent addition of AlGaAs lasers using microlenses and
diffractive coupling," {\em Appl. Phys. Lett.} {\bf 52}(21),
1771--1773.
\item[] Li, R. \& Erneux, T. [1993] "Stability conditions for coupled
lasers: series coupling versus parallel coupling," {\em Optics
Commun.} {\bf 99}(3--4), 196--200.
\item[] Martin--Regalado, J., van Tartwijk, G.H.M., Balle, S.
\& San Miguel, M. [1996]
"Mode control and pattern stabilization in broad--area lasers
by optical feedback,"
{\em Phys. Rev. A} {\bf 54}(6), 5386--5393.
\item[] Roy, R. \& Thornburg, K.S. [1994]
"Experimental synchonization of chaotic lasers,"
{\em Phys. Rev. Lett.} {\bf 72}(13) 2009--2012.
\item[] Sacher, J., Baums, D., Panknin, P., Els\"asser, W. \&
G\"obel, E.O. [1992] 
"Intensity instabilities of semiconductor lasers under current
modulation, external light injection, and delayed feedback,"
{\em Phys. Rev. A} {\bf 45}(3), 1893--1905.
\item[] Sanders, S., Waarts, R., Nam, D., Welch, D.,
Scifres, D., Ehlert, J.C., Cassarly, W.J., Finlan J.M. \&
Flood K.M. [1994] "High power coherent two--dimensional semiconductor
laser array," {\em Appl. Phys. Lett.} {\bf 64}(12), 1478--1480.
\item[] Welch, D.F., Chan, B., Streifer, W. \& Scifres, D.R.
[1988] "High--power, 8W cw, single--quantum--well laser diode
array," {\em Electron. Lett.} {\bf 24}(2), 113--115.
\item[] Winful, H.G. \& Rahman L. [1990]
"Synchronized chaos and spatiotemporal chaos in arrays of coupled
lasers," {\em Phys. Rev. Lett.} {\bf 65}(13), 1575--1578.
\item[] Winful, H.G. \& Wang S.S. [1988]
"Stability of phase locking in coupled seminconductor laser arrays,"
{\em Appl. Phys. Lett.} {\bf 53}(20), 1894--1896.
\end{itemize}

\end{document}